# Dbias: Detecting biases and ensuring Fairness in news articles

Shaina Raza[1*], Deepak John Reji[2], Chen Ding[3]

**Abstract**

Because of the increasing use of data-centric systems and algorithms in machine learning, the topic of fairness is receiving a lot of attention in the academic and broader literature. This paper introduces Dbias (https://pypi.org/project/Dbias/), an open-source Python package for ensuring fairness in news articles. Dbias can take any text to determine if it is biased. Then, it detects biased words in the text, masks them, and suggests a set of sentences with new words that are bias-free or at least less biased. We conduct extensive experiments to assess the performance of Dbias. To see how well our approach works, we compare it to the existing fairness models. We also test the individual components of Dbias to see how effective they are. The experimental results show that Dbias outperforms all the baselines in terms of accuracy and fairness. We make this package (Dbias) as publicly available for the developers and practitioners to mitigate biases in textual data (such as news articles), as well as to encourage extension of this work.

***Keywords:*** *Bias; fairness; Transformer-based models; deep learning; entity recognition; classification; masking*

## 1 Introduction

Natural language processing (NLP) is a branch of artificial intelligence (AI) that assists computers in understanding, interpreting, and manipulating human language [1]. In recent years, the deep learning techniques have demonstrated promising results in narrowing down the gap between sequence-to-sequence learning approaches and human-like performance in a variety of NLP tasks. One common thing about these ML models is that they are often trained on a large text corpus (e.g., Google's BERT, Facebook's BART, and other industry driven products), which may introduce substantial biases into the models. These biases are usually passed on to the models during the training time [2], and often the developers are unaware of these biases.

✉ Shaina Raza Shaina.raza@utoronto.ca
Deepak John Reji deepak.reji@erm.com
Chen Ding cding@ryerson.ca

[1] University of Toronto, Ontario, Canada
[2] Environmental Resources Management, Bangalore, India
[3] Ryerson University, Toronto, Canada

There are numerous examples of biases in ML models due to data, such as Amazon's hiring algorithm[1] that favored men, Facebook's targeted housing advertising that discriminated on the basis of race and color[2], and a healthcare algorithm [3] that exhibited significant racial biases in its recommendations. Data-heavy systems, such as news recommender systems, are also trained on huge volumes of data with limited control over the quality of the training data [4]. These news recommenders often inherit biases from the data, potentially influencing the beliefs and behaviors of news consumers [5].

To mitigate biases, the biased models are usually eliminated, as was the case with Amazon's hiring algorithm, which is not always a feasible solution. As discussed in the literature [6], [7], it is necessary to eliminate these biases early in the data collection process, before they enter the system and are reinforced by model predictions, resulting in biases in the model's decisions. In this study, we aim to eliminate biases in the original data as soon as possible, such as during the data ingestion time.

Bias detection and mitigation are hot topics in academia and industry. Fairness has been defined in a variety of ways by the academic researchers.

[1] us-amazon-com-jobs-automation.
[2] facebook-ads-housing-discrimination-charges

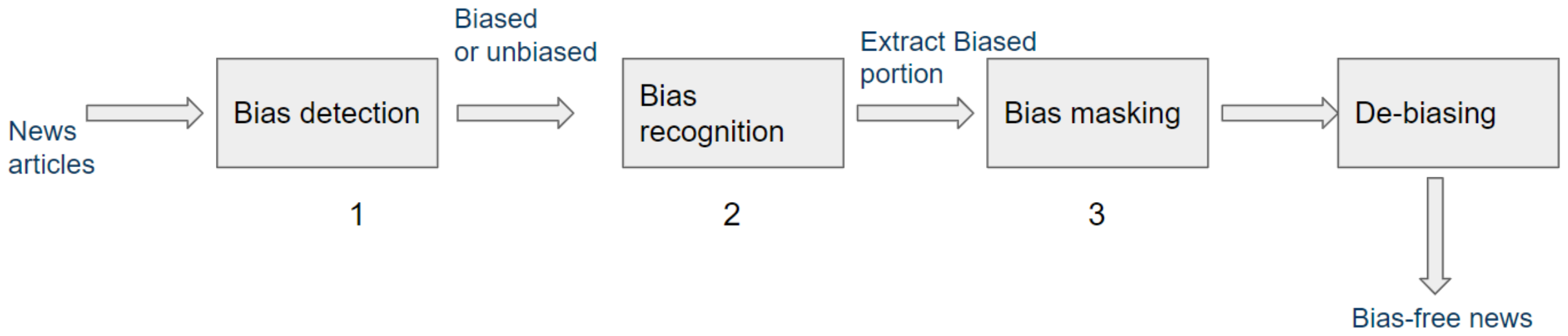

**Figure 1:** Debias – a fair ML pipeline for news articles

Traditionally, the bias is defined as prejudice in favour of or against a particular thing, person, or group in comparison to another, usually in an unjust manner [8], [9]. For example, gender, race, demographic or sexual orientation are some of the examples of the biases. The objective of fairness is to identify and reduce the impacts of various biases [2]. The goal is to prepare these ML systems to avoid perpetuating human and societal biases or adding new biases into the context.

Biases in the ML models can be mitigated at three stages: early, mid, and late [2], [10]. The early stage would be to reduce bias by manipulating the training data before training the algorithm. The mid stage would be to de-bias the model itself, which is also framed as an optimization problem. The late stage refers to reducing the biases by manipulating the output predictions after training the algorithm. Prior research [11] has demonstrated that missing the chance to detect bias at an early stage of the ML pipeline might make algorithmic fairness (mid or late stages of bias mitigation) difficult to achieve. In this paper, we propose a fair ML pipeline that takes raw data and de-biases it early on, ensuring fairness throughout all the phases in the pipeline.

A ML pipeline is composed of several steps that facilitate the automation of ML workflows. Many real-world applications, for example, the Netflix recommender system, Spark healthcare, and Uber forecasting are often represented as ML pipelines. Usually, these pipelines are designed to input some data as features and generate a score that predicts, classifies, or recommends future outcomes. In contrast to typical product-driven ML pipelines, we propose a fair ML pipeline, designed exclusively for mitigating data biases and leveraging fairness in applications.

We develop Dbias – a fair ML pipeline, shown in Figure 1, which mitigates biases in the data and propagates fairness through different phases (data preprocessing, model training, analysis and development) of the pipeline.

As shown in Figure 1, Dbias first detects biases in the text, then it recognizes the biased words, masks and replaces those biased words with new (non-biased or at least less biased) words to de-bias the text. Each of these phases in Dbias pipeline consists of its own training and inference steps (shown in Figure 2).

To the best of our knowledge, there is no ML pipeline developed exclusively to mitigate biases in data. The majority of research on fairness in ML has focused on classification tasks [12], [13]. By focusing exclusively on the fairness of classifiers (e.g., Decision Tree, Logistic Regression), we miss the impact of fairness on other stages in a standard ML pipeline.

FairML [14], FairTest [15], Themis-ml [16] and AIF360 [8] are some of the well-known fair ML pipelines that implement and evaluate the state-of-the-art fairness algorithms. This means that these pipelines use and evaluate off-the-shelf fair ML models, as mentioned in the literature [2] too. In comparison to these approaches (fairness methods and pipelines), Dbias is a self-contained fair ML pipeline that has its own algorithms for detecting and mitigating biases. Dbias ensures that the fairness is maintained throughout the pipeline.

According to research [4], the news media can be biased, and this biased coverage has the potential to significantly influence public perceptions of the reported topics. Biased news media can also result in "filter bubbles" or "echo chambers" [4], which can lead to a lack of understanding of specific concerns as well as a narrow and one-sided point of view. This inspires us to train Dbias to mitigate news media biases.

***Contributions***: We summarize our contributions as:

1. We develop a fair ML pipeline, which we name as Dbias (de-biasing) that de-biases the text (e.g., news text). To make it easier for practitioners to use, we follow the widely acknowledged ML pipeline structure [17] to build Dbias. We develop and package different algorithms for bias detection, recognition, masking and de-biasing into the Dbias pipeline.

2. We make Dbias available as an open-source package distributed under the MIT[3] License. It is publicly shared in the GitHub repository[4] and as the PyPi project[5]. The released package also includes introductory tutorials to the concepts, as well as documentation, usage and guidance to assist data scientists and practitioners in incorporating this package into their work products.
3. Focusing on news media biases in this work, we demonstrate in our GitHub tutorials how we can use Dbias in conjunction with API code blocks (such as those found in Google News API) to reduce biases in news articles.
4. Dbias is released as a generalizable fair ML pipeline, it can be applied to any types of text data. The only requisite is to train the data on the specific domain (e.g., fake news, entertainment, or alike).
5. We conduct extensive experiments to compare the performance of Dbias to that of other cutting-edge fairness methods. We also evaluate the individual components of Dbias to see how well they perform in various experimental settings.

The rest of the paper is organized as follows: Section 2 is the related work, section 3 is the working of the Dbias architecture, Section 4 is about the experimental setup, Section 5 shows and analyzes the results, Section 6 is the discussion, and finally Section 7 gives the conclusion.

## 2 Literature Review

As AI is increasingly used in highly sensitive domains such as news, health care, hiring, journalism, and criminal justice, there is a growing awareness of the consequences of embedded biases and unfairness. Numerous studies have shown that AI is capable of embedding and deploying human and societal biases into the solutions [6], [8]. For example, the Correctional Offender Management Profiling for Alternative Sanctions (COMPAS) algorithm mislabeled African-American defendants nearly twice as often as white defendants[6]. The inability of ML models to mitigate these undesirable biases is a significant impediment to AI reaching its full potential.

Fairness [2] is a multi-faceted concept that varies by culture and context. It is quite difficult to have a standard definition of fairness as each definition depends on a different use case and organization. Distinct definitions of fairness can lead to different outcomes. There exists at least 21 mathematical definitions for fairness in politics [18]. A decision tree[7] on different definitions of fairness is provided by the University of Chicago. Overall, it is quite difficult to meet many definitions of fairness at the same time.

Most of the definitions on fairness focus on either individual fairness (treating similar individuals fairly) or group fairness (equitably distributing the model's predictions or outcomes across groups) [2], [10]. Individual fairness aims to ensure that statistical measures of outcomes are equal for people who are statistically similar. Group fairness divides a population into distinct groups based on protected characteristics, with the goal of ensuring that statistical measures of outcomes are comparable across groups.

Bias is defined as an inclination or prejudice for or against one person or group, especially in an unfair manner [6]. While algorithmic bias is frequently discussed in the literature of ML, in most situations it is the underlying data that introduces bias. For example, if there had been an equal amount of data for men and women in Amazon's recruiting algorithm, the algorithm might not have biased as much.

***Fairness algorithms***: In the research of AI and ML fairness [2], [8], [10], the bias mitigation algorithms are categorized into three broad types: (1) pre-processing algorithms; (2) in-processing algorithms; and (3) post-processing algorithms. These algorithms are briefly discussed below.

*Pre-processing algorithms*: The pre-processing algorithms attempt to learn a new representation of data by removing the information associated with the sensitive attribute, while retaining as much of the actual data as possible. This technique manipulates the training data prior to training the algorithm. Well-known pre-processing algorithms are:

- A reweighting [19] algorithm that generates weights for the training samples in each (group,

[3] https://opensource.org/licenses/MIT

[4] https://github.com/dreji18/Fairness-in-AI

[5] https://pypi.org/project/Dbias/

[6] machine-bias-risk-assessments-in-criminal-sentencing.

[7] http://aequitas.dssg.io/static/images/metrictree.png

label), without changing the actual feature or label values.

- The learning fair representations [20] algorithm that discovers a latent representation by encoding the data, while concealing information about protected attributes. Protected attributes are those attributes that divide a population into groups whose outcomes should be comparable (such as race, gender, caste, and religion) [2].
- The disparate impact remover [12] algorithm modifies feature values to improve group fairness, while keeping rank ordering within groups.
- Optimized pre-processing [21] algorithm learns a probabilistic transformation that edits data features and labels for individual and group fairness.

Usually, pre-processing techniques are easy to use as the modified data can be used for any downstream tasks without requiring any changes to the model itself.

*In-processing algorithms*: In-processing algorithms penalize the undesired biases from the model, to incorporate fairness into the model. The in-processing technique influences the loss function during the model training to mitigate biases. In the past, the in-processing algorithms [22], [23] have been used to provide equal access to racially and ethnically diverse group. Some of the example in-processing algorithms are listed below:

- Prejudice remover [22] augments the learning objective with a discrimination-aware regularization term.
- Adversarial De-biasing [24] algorithm learns a classifier to maximize the prediction accuracy while decreasing an adversary's ability to deduce the protected attribute from the predictions.
- Exponentiated gradient reduction [25] algorithm breaks down fair classification into a series of cost-sensitive classification problems, returning a randomized classifier with the lowest empirical error subject to fair classification constraints.
- Meta fair classifier [23] algorithm inputs the fairness metric and returns a classifier that is optimized for the metric.

The in-processing technique is model or task-specific, and it requires changes within the algorithm, which is not always a feasible option.

*Post-processing algorithms:* The post-processing algorithms manipulate output predictions after training to reduce bias. These algorithms can be applied without retraining the existing classifiers (as in in-processing). Some of the post-processing algorithms are:

- Reject option classification [26] algorithm gives favorable outcomes (labels that provide advantage to an individual or group, e.g., being hired for a job, not fired) to unprivileged groups and unfavorable outcomes (not hired for a job, fired) to privileged groups (has historically been at a systemic advantage).
- Equalized odds [27] algorithm changes the output labels to optimize equalized odds through linear programming.
- Calibrated equalized [28] odds optimizes score outputs to find probabilities with which to change output labels with an equalized odds objective.

Usually, the post-processing technique requires access to protected attributes late in the pipeline.

Recently, the software engineering community has also started to work on fairness in ML, specifically fairness testing [29]. Some work has been done to develop automated tools, such as AI Fairness 360 [8], FairML [14] , Aequitas [30], Themis-ML [16], FairTest [15], that follows a software development lifecycle.

***Transfer learning techniques:*** Transfer learning is a technique to transfer the knowledge contained in larger, different but related source domain to a target domain [31]. The goal is to improve the performance of target domain with the existing knowledge of the source domain. Bidirectional Encoder Representations from Transformers (BERT) [32] is an example, which has shown state-of-the-art performance in many tasks like classification, question answering, and so on.

Li et al. 2021 [33] study the gender bias inside the Transformer-based model (BERT). They calculate the attention scores for the corresponding gender pronouns and occupations, swap the gender pronouns to eliminate the position effect on bias judgement, and then again check the consistency of the gender bias associated with the occupation. Sinha and Dasgupta [34] employ the BERT model to detect biases from the text. Both models, while equally important, are primarily concerned with the detection and extraction of biased sentences.

The task of identifying a named entity (a real-world object or concept) in unstructured text and then classifying the entity into a standard category is known as named entity recognition [35]. Mehrabi et al. 2020 [13] use named entities to determine whether female names are more frequently tagged as non-person than male names. Some other researchers [36], [37] use the named entities to identify biases based on occupation, race, and demographics. These named entity recognition models usually recognize the biased entities from the data. However, the mitigation technique is not their objective.

The masked language modeling is also used to identify biases. Based on two crowdsourced datasets, Kaneko and Bollegala 2021 [38] propose a technique to accurately predict different types of social biases in text. They [38] demonstrate that social biases do exist in masked language models and suggest developing methods to robustly debias pre-trained masked language models as a future direction.

***Fairness toolkits:*** We discuss the fairness toolkits here:

- FairML [14] is a toolkit that uses a few ranking algorithms to quantify the relative effects of various inputs on a model's predictions, which can be used to assess the fairness in models.
- FairTest [15] is a Python package that learns a special decision tree that divides a user population into smaller subgroups with the greatest possible association between protected features (e.g., gender, race, and other features deemed sensitive) and algorithm outputs.
- Themis-ml [16] is a Python library that implements several state-of-the-art fairness-aware methods that comply with the sklearn API.
- AIF360 [8] consists of a number of fairness metrics for datasets and state-of-the-art ML models to mitigate biases in the datasets and models.

***Fairness metrics****:* To ensure the fairness, the ML community has also proposed various fairness metrics. For example, the disparate impact ratio compares the rate at which an underprivileged groups (groups having systematic biases e.g., females) receives a particular outcome compared to a privileged group (groups with systematic advantages, e.g., males) [12]. In some works [2], the number of positives and negatives for each individual or group, difference of means, odds ratio are also computed to measure fairness.

***Comparison with the state-of-the-art approaches:*** While each of the previous works discussed in this section is valuable and incremental, they are primarily concerned with fairness across different tasks (pre-processing, in-processing, and post-processing). These models either remove biases or ensure fairness during pre-processing, in-processing or post-processing. The fairness toolkits (AIF360, FairML, FairTest, etc.) also take the existing approaches to mitigate the biases in different groups (gender, populations, religions).

According to the research [10], the existing works for ensuring fairness and mitigating biases fall into four main domains: ML, information retrieval, recommender systems and human computer interaction. A number of related works [10] are also focusing on the explainability approaches that contribute to the transparency as well as the perception of fairness, which is a potential future direction.

In this work we combine the strength of deep neural networks and Transformer-based architecture into our fair ML pipeline. Our work falls into the information retrieval and ML categories to ensure fairness. We propose that fairness can be achieved by ensuring that each component of the ML pipeline is fair. We do not rely on existing built-in fairness models to ensure that data or model is fair; rather, we construct a fair ML pipeline comprised of multiple components that takes raw data and de-biases it. Also, compared to the previous methods, we consolidate various bias mitigation methods, such as bias detection and mitigation in a single pipeline. Dbias is developed with reusability and generalizability in mind. The only requisite is to fine-tune the model on the domain-specific data.

## 3 DBias – a fair ML pipeline

In this section, we define the preliminaries, problem definition, overview, and underlying working of Dbias.

### 3.1 Preliminaries

We use the following key terms [2], [6] in this work.

- *Bias* is a type of systemic error.
- A *protected attribute* divides a population into groups, for example, race, gender, caste, and religion.
- A *privileged* value of a protected attribute denotes a group that historically has a systematic advantage, for example, male gender, white race.
- An *underprivileged* group faces prejudice based on systematic biases such as gender, race, age, disability, and so on.

- *Group fairness* means that the protected attribute receives similar treatments or outcomes.
- *Individual fairness* means that similar individuals receive similar treatments or outcomes.
- *Equalized odds* [39] is a statistical notion of fairness that ensures classification algorithms do not discriminate against protected groups.
- *Fairness* or *fair* generally refers to the process of undoing the effects of various biases in the data and algorithms.

The goal of *fairness* is to mitigate the unwanted biases that benefit the privileged groups and disadvantage the unprivileged. In Table 1, we define some of the biases in news media [40], [41] that we try to mitigate in this work. However, depending on the data on which our model is fine-tuned, we can also address many other types of biases.

**Table 1**: Some biases in news domain

| **Bias** | **Example** |
|---|---|
| Gender | All man hours in his area of responsibility must be approved. |
| Age | Apply if you are a recent graduate. |
| Racial/ Ethnicity | Police are looking for any black males who may be involved in this case. |
| Disability | Genuine concern for the elderly and handicapped. |
| Mental health | Any experience working with retarded people is required for this job. |
| Other biases addressed: Religion, education, political ideology (liberal, conservative) | |

## 3.2 Problem definition

Given a set of news articles that may contain a variety of biases, the task is to detect, recognize and undo those biases from the data. The goal of this research is to mitigate unfairness in the news domain.

## 3.3 Overview of Dbias

In this section, we present the workflow of Dbias in Figure 2. We use the word fair ML "pipeline" to refer to multiple sequential steps from data extraction to preprocessing to modeling, data transformation to deployment. The main phases in this pipeline are bias detection, bias recognition, bias masking and de-biasing. Each phase gets the input from the preceding phase, except for the first phase (bias detection phase) that gets the input from a news API (e.g., Google news). Each phase has a training, validation, and testing task. The transformation of the data from one phase to another phase is shown by a trapezoid.

Each model is served as an inference API and can function individually as well. The output from each trapezoid goes to the respective inference API. For example, the transformed data from the bias detection phase goes into the bias detection inference API. Once the models are trained and tested, each of them is registered and deployed so that it can be used as a callable API. The goal of the inference API is to execute inference queries over the new transformed data without the need of loading all the heavy weight python libraries. The final output of the pipeline is the news articles (or any text) that are free from biases. There are also various stages in the pipeline where we evaluate the modules using fairness metrics (not pictured). These will be discussed in the evaluation section. Next, we explain each phase of the Dbias pipeline.

### 3.3.1 Data collection and preparation phase

We prepare the annotated dataset MBIC (Media Bias Including Characteristics) [41] that represents various bias instances in the news articles to train our models. More details about the dataset are given in Section 4.

### 3.3.2 Model training, fine-tuning, and testing

In this work, we fine-tune the state-of-the-art pretrained Transformer models, such as BERT, DistilBERT, RoBERTa on our MBIC dataset for various downstream tasks. We use the training checkpoints of these models (BERT, DistilBERT and RoBERTa) and fine-tune them using MBIC dataset. For example, we fine-tune the DistilBERT model on the MBIC dataset for the bias classification task. Similarly, we fine-tune the RoBERTa on our MBIC dataset for the bias recognition task. We also adapt the BERT architecture for masking the biased words and filling in those masked words with the non-biased words. We discuss the details of each of such tasks in Section 3.4.

### 3.3.3 Model registering and sharing

We have registered and shared our models as TensorFlow checkpoints on the Huggingface.co website. We create the model cards for the tasks: bias detection[8] and recognition[9]. Model cards are the markdown files providing useful information for reproducibility and sharing the model. We also provide the inference (running live data points) API along with each model card. Our purpose is to make it easier for other researchers and developers to use our model and to contribute.

[8] https://huggingface.co/d4data/en_pipeline

[9] https://huggingface.co/d4data/bias-detection-model

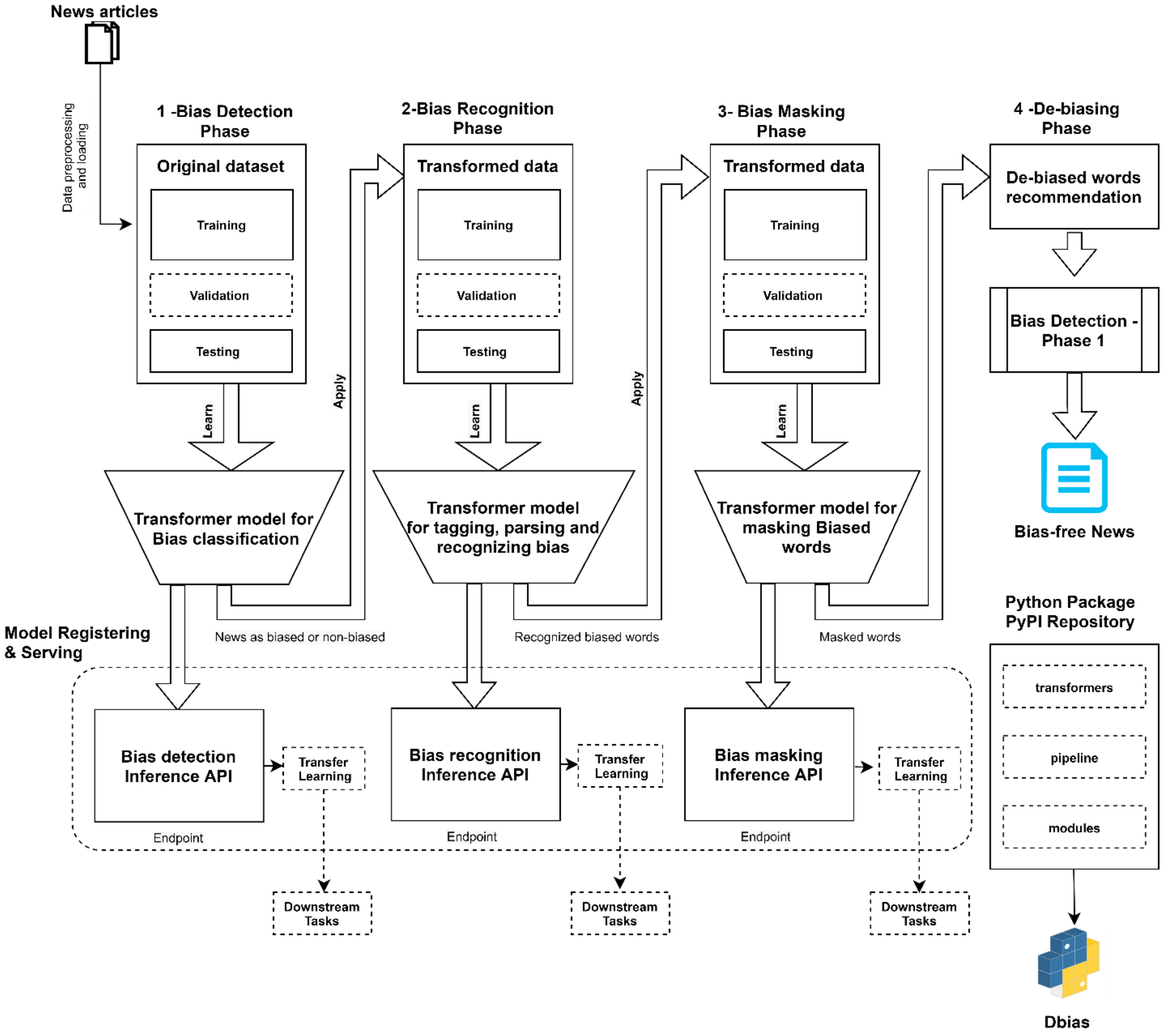

**Figure 2:** Dbias and its debiasing workflow

### 3.3.4 Packaging

We group different modules of Dbias pipeline into a single package. Our whole package is hosted on http://pypi.org under the MIT license and can be installed using the command *pip install Dbias*. Next, we discuss our methodology.

## 3.4 Methodology

The specific tasks to perform in this work are:

- Bias detection: To detect whether a news article is biased or not.
- Bias recognition: To recognize the biased words or phrases from the news articles.
- De-biasing: To de-bias the data by replacing the biased words or phrases from the news article with unbiased or at least less biased word(s). The de-biasing also consists of masking, i.e., to hide the biased words.

### 3.4.1 Bias detection module

The input to Dbias is a set of news articles that potentially contain biased sentences. The task of the bias detection module is to predict if a sentence is biased or non-biased.

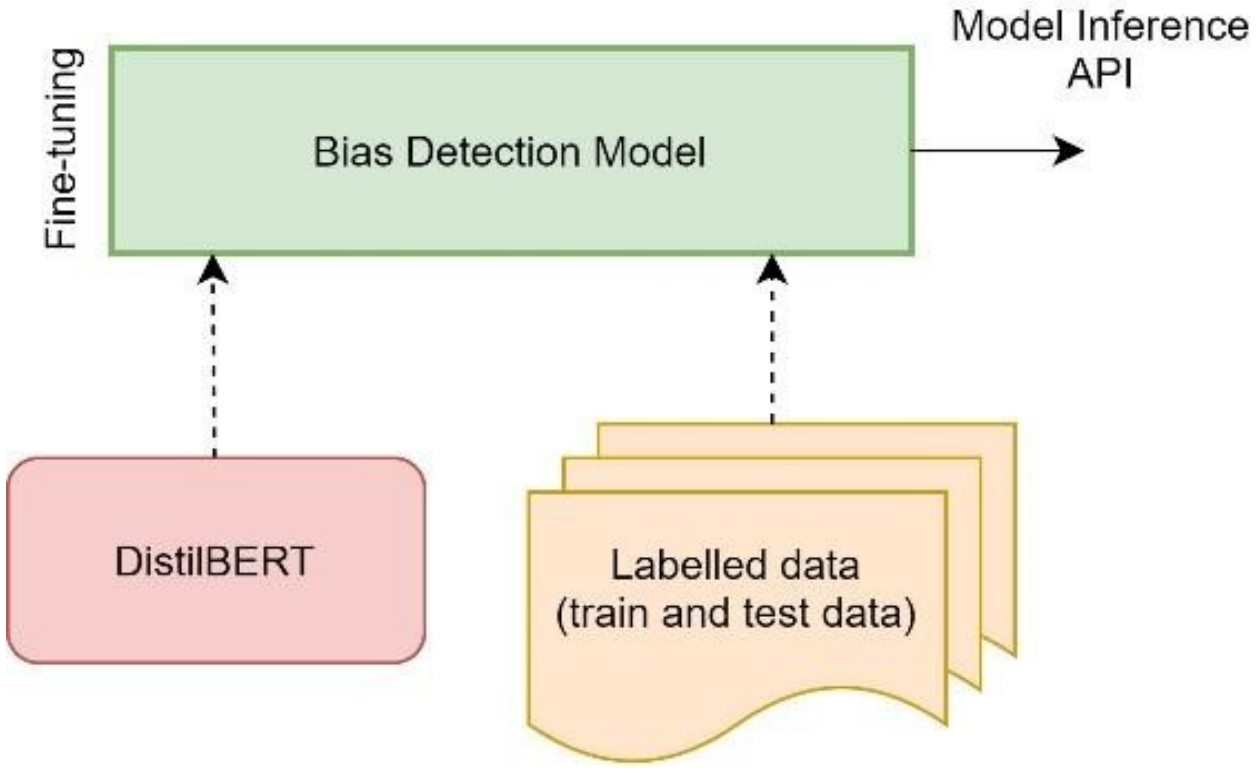

**Figure 3:** Bias detection module

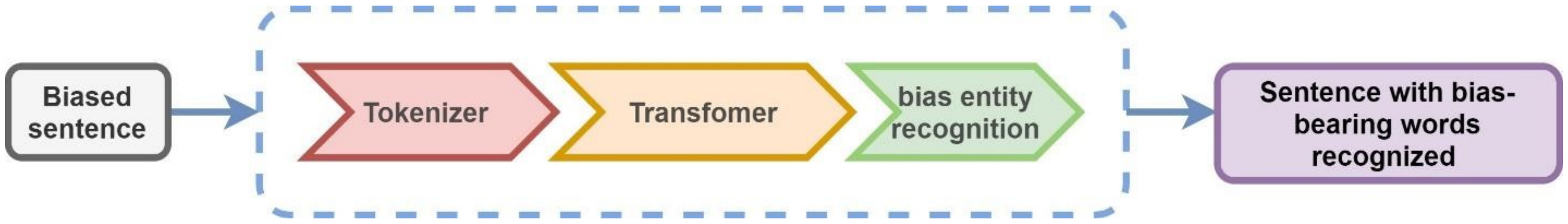

**Figure 4:** Bias recognition pipeline

For instance, given the news headline, "Don't buy the pseudo-scientific hype about tornadoes and climate change"[10], the bias detection module should be able to detect the sentence's bias and classify it as biased news.

In our preliminary experiments, we fine-tuned and evaluated various Transformer-based models (e.g., BERT, GPT-2, and others from the HuggingFace's Transformers library) on our dataset and found that the DistilBERT [42] model achieves both higher accuracy and faster inference speed, which is why we chose to work with it. DistilBERT is a distilled (approximate, faster and smaller) version of BERT.

For binary classification, we use binary-cross entropy loss with sigmoid activation function. The output from the bias detection model is a set of news articles that are classified as biased or non-biased. Our bias detection module is shown in Figure 3.

### 3.4.2 Biased recognition module

The second module in the Dbias pipeline is the bias recognition module. The task of the bias recognition module is to annotate the biased words in the news articles with tags, indicating that each tagged word is biased. This module takes as input a set of news articles that have been identified as biased in the preceding module (bias detection), and outputs a set of news articles where the biased words are picked and recognized. The news headline "Don't buy the pseudo-scientific hype about tornadoes and climate change", for example, has already been classified as a biased sentence by the preceding bias detection module, and the biased recognition module can now identify the term "pseudo-scientific hype" as a biased word.

Traditionally, named entity recognition (NER) is the task of identifying a named entity (a real-world object or concept) in unstructured text and then classifying the entity into a standard category [35]. Though NER is not directly related to bias identification, the research [6], [36] shows that NER models can be used to examine the existence and level of biasness in data, which obviously helps in mitigating biases in various applications. For example, an NER model can be used to find if there are more female names tagged as non-person than male names [13] or to identify biases based on occupation, race, and demographics [36].

In our work, we take a unique approach to identifying biases in news articles. Rather than looking for conventional entities (Name, Location, Event, and Organization, which are primarily nouns), we look for bias-bearing words associated with each entity. Specifically, we refer to each entity in the NER task as a bias-bearing entity that is manifested in syntax, semantics or in the linguistic context (e.g., the word 'pseudo-scientific hype' is a biased word).

We show our bias recognition module, which is also a pipeline, in Figure 4.

We use a Transformer-based model in our bias recognition module. A standard NER [35] task has traditionally been viewed as a sequence labelling problem with word-level tags. The standard NER models are based on dictionaries or rules that may fail to recognize references to unknown entities in the text (for example, biased words as in our work). Therefore, we employ a Transformer-based model as a wrapper for the bias recognition task. The Transformer-based model can also detect long-term context in data, allowing us to identify more bias-bearing words in the text.

We employ the RoBERTa [43], a retrained version of BERT with enhanced training methodology, and fine-tune it on our dataset. By including the Transformer in the standard NER pipeline, our bias recognition model can now identify many other biased words beyond the ones defined in the dataset. The final output from the bias recognition module is a set of news articles, where the biased words have been identified.

### 3.4.3 De-biasing module

The de-biasing module is the most important part of Dbias. The purpose of this entire workflow is to reduce the bias effects of news text while maintaining its semantic content, which has a wide range of applications in many domains as well. For example, we could convert the biased news headline "Don't buy the *pseudo-scientific hype* about tornadoes and climate change" to a non-biased sentence as "Don't buy the *information* about tornadoes and climate change". Our third module, de-biasing module is specifically designed to accomplish this goal.

[10] https://nypost.com/2021/12/12/dont-buy-the-psuedo-scientific-hype-about-tornadoes-climate-change

The input to the de-biasing module is a collection of news articles from the bias recognition module with identified biased words, and the output is a list of recommendations for each news article with non-biased or at least less biased words.

Our de-biasing approach comprises of two stages: Bias Masking and Fairness Infill stages. In the bias masking stage, we mask the position of each biased word (token) within each news article. In the fairness infill stage, we fill the masked token positions with new words according to its context in the sentence.

***Bias Masking***: Our approach to masking is different from the Masked Language Modeling (MLM) task of the standard BERT model. Typically, the MLM task takes a sentence, and randomly masks 15% of the words in the input and then run the entire masked sentence through the model to predict the masked words. In our work, we only mask words that have been flagged as biased by the previous bias recognition module. We demonstrate our approach to bias masking through an example from the news headline[11] in Figure 5.

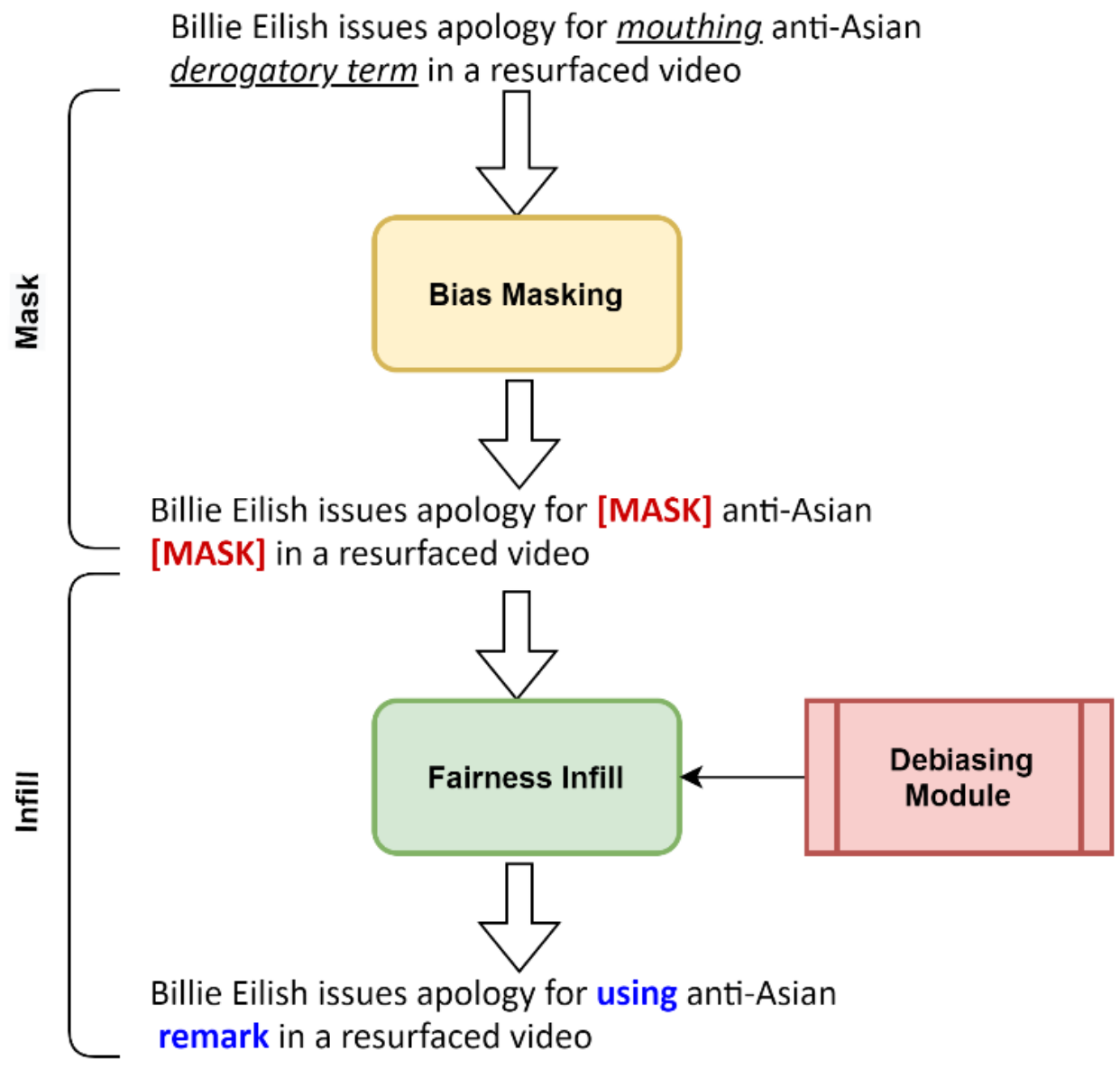

**Figure 5:** De-biasing example with MLM

[11] https://www.bbc.com/news/newsbeat-57564878

As can be seen in Figure 5, we have provided two masks in the news headline. Both of these words have been identified as consisting of biases by the bias recognition module. This approach is different from a typical MLM task [32] where only one token is masked and infilled at a time.

***Fairness Infilling***: We propose a unique mask shifting technique that can mask and unmask more than one token at a time in a sentence. For example, our mask shifting technique can breakdown the above news headline (Figure 5) to two instances (based on two token masks) as below:

- "Billie Eilish issues apology for [MASK] an anti-Asian derogatory term in a resurfaced video."
- "Billie Eilish issues apology for mouthing an anti-Asian [MASK] in a resurfaced video."

Then, both instances will be processed sequentially via mask shifting technique, and each masked token will be filled one at a time before the final sentence is constructed.

Our proposed Fairness Infilling stage can be considered as generalizing the cloze task (Wu et al. 2019) from single tokens to spans of tokens with unknown length. Our assumption behind this fairness filling is that the new words that are filled in are less or non-biased, which has been validated through our demonstration and experiments.

*Recommending Words*: We recommend a couple of substitute tokens that can be used to infill for each masked token during the Fairness Infilling stage. For example, we can recommend top-k (k=5,10 or so) substitutes for each masked token in the de-biased sentence. We send the top-k recommended sentences (infilled with new tokens) again to the bias detection model (first module) to see the probability of biasness. If the probability of biasness is less than 0.5 or less than the probability of the previous de-biased sentence, we output the sentence as the final output.

## 4 Experimental setup

Fairness is a complex concept with no one-size-fits-all solution. Considering various definitions [10] and solutions of fairness [8], it is not possible to find one benchmark solution for mitigating biases. However, it is still important to evaluate the working of our Dbias pipeline. We evaluate our architecture with two goals in mind: (1) to demonstrate capabilities of our package in terms of bias detection and mitigation algorithms, and (2) to demonstrate how a user can understand the behavior of bias mitigation algorithms on the dataset and make an appropriate choice based on the business need.

## 4.1 Dataset

In this work, we primarily use MBIC – A Media Bias Annotation Dataset [41]. MBIC provides the 17,000 annotated sentences from approximately 1,000 news articles belonging to different news sources (HuffPost, MSNBC, AlterNet, Fox News, Breitbart, USA Today and Reuters and other). The dataset has around 10k biased and 7k unbiased labels. The data collection focuses on news articles depicting racial, gender, religious, political, age and related biases, some of these biases are also mentioned in Table 1.

In the original dataset, these biases are identified through crowdsourcing, where the crowdsourced annotators picked the bias-bearing words from the news text. We also calculated the biases (gender, race, ethnicity, educational, religion, and language) from the sentences, which are different from the annotators' labeling on gender, age, and education. While MBIC dataset itself has labels created by crowdsource workers. We obtained an additional list of most commonly used biases from this source [44] and use it to annotate the dataset further. We went through all the records one by one and added the biased words in the words list. The goal is to have a populated list of biased words from the text that will be later used to train the model.

The dataset features used in this work are:

- *Sentence:* Sentence from the news article
- *News Link:* URL of the news from which the news is taken
- *News Outlet:* News sources (USA today, MSNBC)
- *Topic*: Topic of news (gun-control, coronavirus, white-nationalism, etc.)
- *Age:* Age of the person annotating the sentence as biased or non-biased
- *Gender*: Gender of the person involved in annotating the data
- *Education:* Education of the annotator
- *Biased words:* some of the biased words added by annotators and some are added by us.

*Label:* News as biased or non-biased

The *age* feature in the dataset is found as the numeric values, however, we categorize it into 3 groups (elder, young, adult). Similarly, we categorize the *education* groups based on degrees (high school, undergraduate, graduate) information.

***Groups***: In this work, we consider the following protected attributes from the dataset: 'gender': [Male, Female], 'age' ['Elder', 'Young', 'Adult'], 'education': ['College degree', 'high school'], language: ['English speaker', 'non-English speaker'], race: ['black', 'white', 'caucasian', 'asian']. We also consider the following privileged attributes: 'Male' (gender), 'College degree' (education), 'English Speaker' (language), 'White' (race). In addition, we consider the following unprivileged attributes: 'Female' (gender), 'high school (education), 'non-English Speaker' (language), 'black' (race), 'asian' (race). We put together these attributes into privileged/ unprivileged groups based on the number of biased words associated with each attribute. We choose these attributes based on the marginality experienced by different groups across societal settings, such as gender, race, ethnicity, religion, disability, and sexual orientation, as reported in the pertinent literature [45].

We show the specific subsets of the data for the main identities (male, female, white, black, etc.) associated with each group in Table 2. The identities are assigned to each group (privileged, unprivileged) based on the count of biased words associated with each identity. We consider an identity as belonging to an unprivileged group if it is associated with group-specific biased words in majority.

**Table 2:** Distribution of identities based on biased words

| Identity | Count |
|---|---|
| Female | 5548 |
| Male | 2494 |
| Black | 4536 |
| White | 1761 |
| English Speaker | 810 |
| Non- English Speaker | 3744 |
| Young | 1301 |
| Older | 3636 |
| College degree | 945 |
| High school | 3573 |

Our preliminary analysis of this dataset indicates that it can be used to capture a variety of biases, particularly because it measures public perceptions of bias. So, we chose this dataset as our primary source of data. Additionally, the articles in this dataset cover a broad range of topics (political, scientific, ethnicity and so), all of which are relevant to our objective of identifying various forms of textual bias. The following are some of the findings from our exploratory analysis of the data.

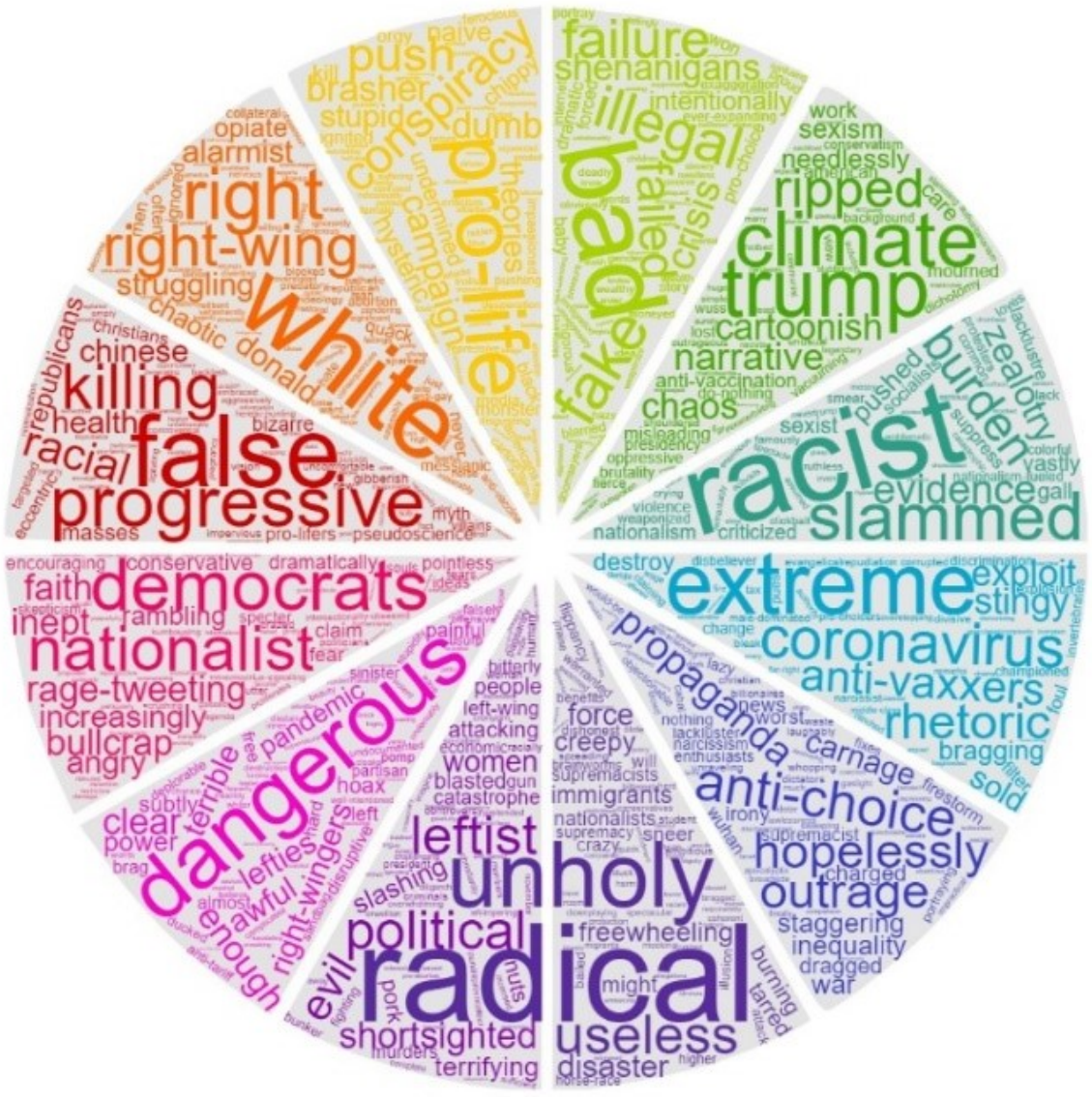

**Figure 6:** Biased words in the news articles

Figure 6 shows a word cloud made up of text from the news articles. Among the notable words are "unholy", "radical", "racist", "slammed", "right-wing", and "rage-tweeting", some of which represent political biases during the 2020 US elections. Figure 7 (a) and (b) shows the top biased bigrams and trigrams respectively from the news articles.

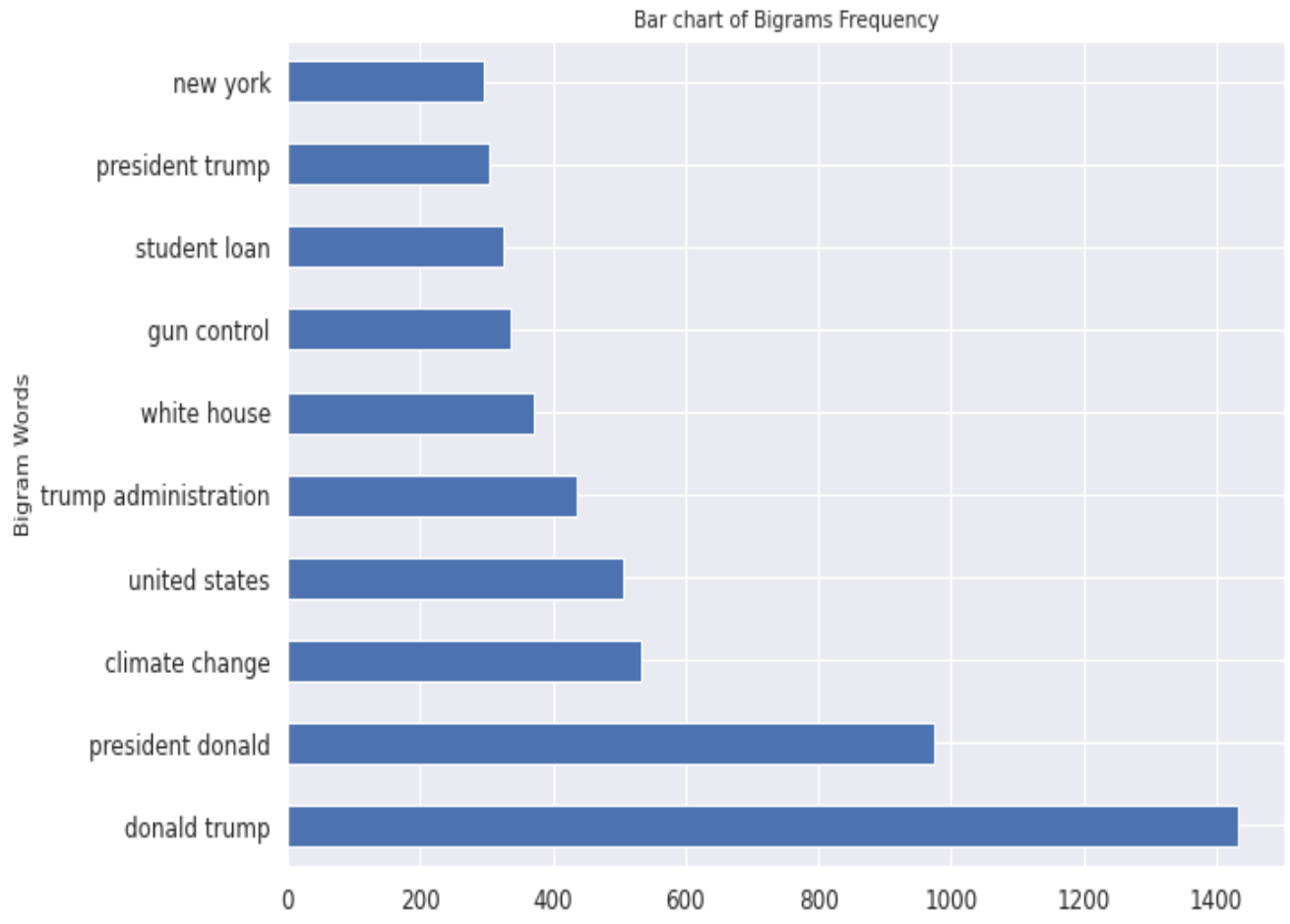

**Figure 7 (a)**: Bigrams of biased words in the news

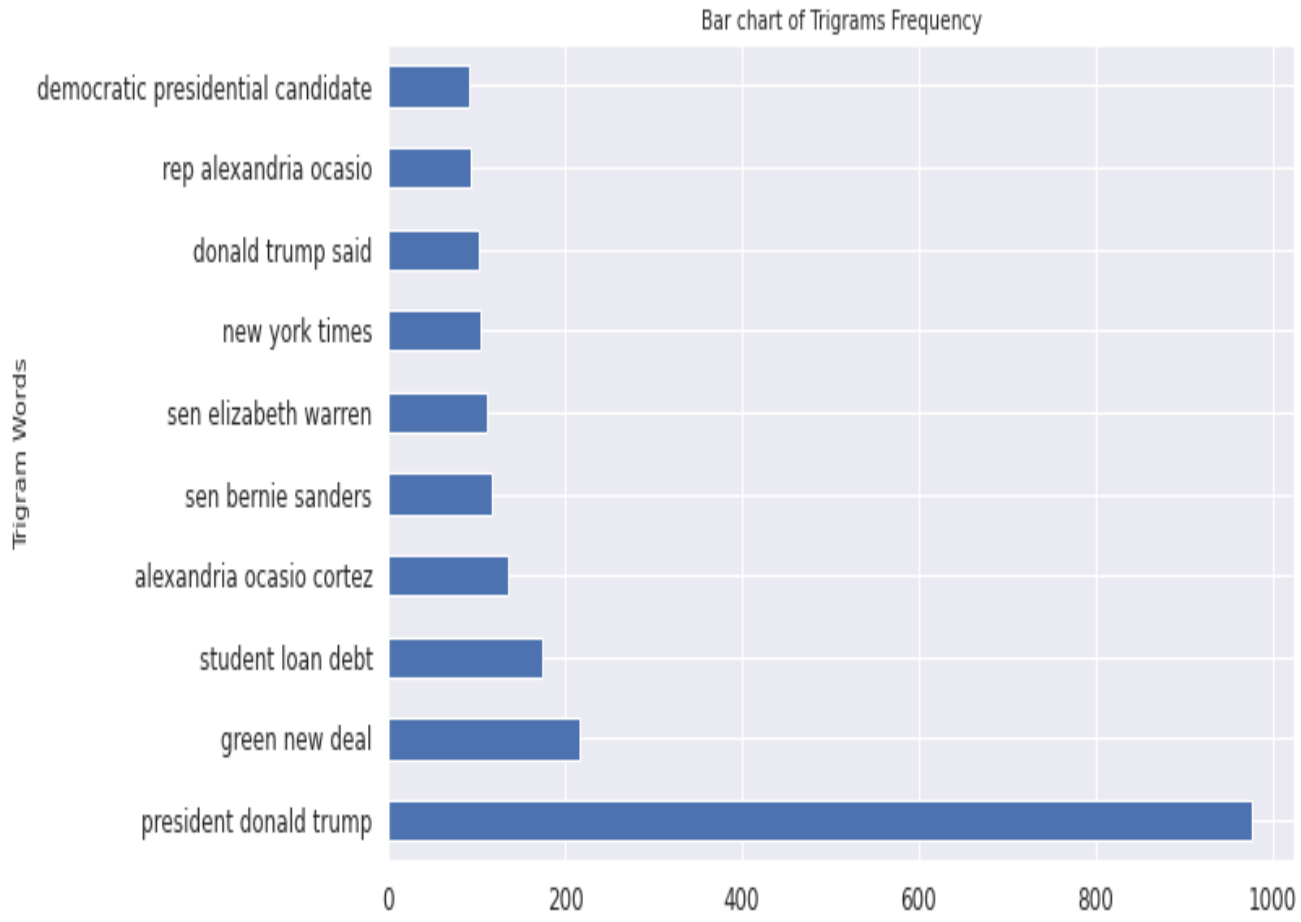

**Figure 7 (b):** Trigrams of biased words in news

Some of the biased words that we see in these news articles are related to Elections 2020, COVID-19, climate change, students' loan and so on.

### 4.2 Baseline Methods

We could not find a single state-of-the-art model that can perform all these tasks: (1) bias detection, (2) bias recognition, (3) de-biasing altogether. Thus, we follow the evaluation strategy in the related work [2], which categories fairness methods into: (1) fairness pre-processing, (2) fairness in-processing, and (3) fairness post-processing methods. We also use other baseline methods to evaluate the effectiveness of individual modules of Dbias, which are detailed in their respective sections. The baseline methods given in Table 3 are the bias mitigation or fairness methods. We use the classification method with each baseline method (Table 3) to detect the existence of the biases.

### 4.3 Evaluation strategy

Our evaluation strategy is given below:

*Evaluating Dbias against state-of-the-art baselines*: The outline of this evaluation is given below:

1. We split the original dataset into training and test sets.
2. We quantify the biases using a fairness evaluation metric on the original dataset (test set) based on different groups (privileged/ unprivileged).
3. We use AutoML[12] to select the best performing classification method for each baseline method to detect and evaluate the existence of the biases.
4. We de-bias the data using each of the baseline's specific methodology.
5. We evaluate the capability of each method on fairness (i.e., how many biases are mitigated by each method) on the transformed data using fairness metrics.
6. Finally, we again test each method to detect the biases in transformed data.

*Evaluating the effectiveness of bias detection module of Dbias:* We evaluate various classification, models with our fine-tuned DistilBERT to determine which model/setting yields the most accurate results.

*Evaluating the effectiveness of bias recognition module of Dbias*: We compare several configurations of different NER models to find which setting gives us the best results.

[12] automl

**Table 3**: Baseline methods

| Model | Description |
|---|---|
| **Fairness Pre-processing models** | |
| **Disparate impact remover** [12] | Disparate Impact Remover is a pre-processing approach for increasing fairness between groups (privileged and unprivileged). This technique edits the feature values (e.g., the features that are privileged, unprivileged) so that the data can be made unbiased while preserving relevant information in the data. Once this algorithm has been implemented, any machine learning or deep learning model can be built using the repaired data. The Disparate Impact metric is then used to validate if the model is unbiased (or within an acceptable threshold). In this baseline method, we use a couple of methods using AutoML and reporting the results with the best performing model. For this baseline, the Logistic Regression gave us the best results. |
| **Reweighing** [19] | Reweighing is a preprocessing technique that weighs the examples in each group (such as privileged, unprivileged groups) to ensure fairness before classification. This algorithm transforms the dataset to have more equity in positive outcomes on the protected attribute(s) for both privileged and unprivileged groups. We run a couple of algorithms on the transformed data and report the result with the best performing model, which is Support Vector Machine (SVM) in this experiment. |
| **Fairness In-processing Models** | |
| **Adversarial Debiasing** [24] | Adversarial debiasing is based on the generative adversarial network (GAN) model. Through training, this model debiases the word and general feature embeddings. This is an in-processing technique that learns the definitions of fairness, such as demographic parity, equality of odds, and quality of opportunity, so that a discriminator (part of GAN) has been tasked with predicting the protected attribute encoded in the bias of the original feature vector, while a competing generator (part of GAN) has been tasked with producing more debiased embeddings to compete with the discriminator. |
| **Exponentiated Gradient Reduction** [25] | Exponentiated gradient reduction is an in-processing technique that reduces the fair classification down into a series of cost-sensitive classification problems. It also returns a randomized classifier with the lowest empirical error (approximation of the expected error), if the fair classification rules are met. |
| **Fairness post-processing models** | |
| **Calibrated Equalized Odds Postprocessing** [28] | Calibrated equalized odds postprocessing is a postprocessing technique that optimizes over calibrated classifier score outputs to find probabilities for changing output labels with an equalized odds objective. |
| **Equalized Odds postprocessing** [27] | Equalized odds postprocessing is a post-processing technique that uses a linear program to find probabilities for changing output labels in order to optimize equalized odds. |
| **Our approach** | |
| **Dbias** | Our approach mitigates biases during the pre-processing stage and ensures that fairness is carried on throughout the ML pipeline to give a fair representation of data. |

*Evaluating the effectiveness of masking technique of Dbias*: We explore the influence of different masking probability to find the best setting for the masking.

### 4.4 Evaluation metrics

To assess the performance of our proposed model, we use accuracy (ACC), precision (PREC), recall (Rec), F1-score (F1), Disparate Impact (DI), and *Generalized Mean of Bias AUCs (G-AUC)* as the evaluation metrics. A confusion matrix determines the information about actual and predicted values, as shown in Table 4.

**Table 4**: Confusion Matrix

| | Actual Fake | Actual Real |
|---|---|---|
| **Predicted Fake** | TP | FP |
| **Predicted Real** | FN | TN |

The variables TP, FP, TN, and FN in the confusion matrix refer to the following:

- *True Positive (TP):* number of biased news that are identified as biased.
- *False Positive (FP):* number of unbiased news that are identified as biased news.
- *True negative (TN):* number of unbiased news that are identified as unbiased news.
- *False negative (FN):* number of biased news that are identified as unbiased news.

For the Prec, Rec, F1 and ACC, we perform the specific calculation as shown in the Equation (1), (2), (3) and (4) respectively:

$$Prec = \frac{TP}{TP + FP} \quad (1)$$

$$Rec = \frac{TP}{TP + FN} \quad (2)$$

$$F1 = \frac{TP}{TP + \frac{1}{2}(FN + FP)} \quad (3)$$

$$ACC = \frac{TP + TN}{TP + TN + FP + FN} \quad (4)$$

*Disparate Impact (DI)* [15] is an evaluation metric to evaluate fairness. It compares the proportion of individuals that receive a positive output for two groups: an unprivileged group and a privileged group. The industry standard for DI is a four-fifths rule [46], which means if the unprivileged group receives a positive outcome less than 80% of their proportion of the privileged group, this is a disparate impact violation. An acceptable threshold should be between 0.8 and 1.25, with 0.8 favoring the privileged group, and 1.25 favoring the unprivileged group [46]. Mathematically, it can be defined as:

$$DI = \frac{\frac{num_positives(privileged=False)}{num_instances(privileged=False)}}{\frac{num_positives(privileged=True)}{num_instances(privileged=True)}} \quad (5)$$

where num_positives is the number of individuals in the group: either privileged=False (unprivileged), or privileged=True (privileged), who received a positive outcome. The num_instances are the total number of individuals in the group.

The DI calculation is the proportion of the unprivileged group that receive the positive outcome divided by the proportion of the privileged group that received the positive outcome. This means DI ratio is the ratio of positive outcomes (Bias=1) in the unprivileged group (females, elderly, non-English speakers, no higher education) divided by the ratio of positive outcomes in the privileged group (males, adults, English speakers, higher education).

Although DI is not specifically designed for analyzing text-based biases, taking inspiration from related works [11], we measure the biases on three specific subsets (number of positives, number of negatives and total number of instances) in the test set that mention the identities (gender, education, spoken language) of specific groups using biased or unbiased words.

*Generalized Mean of Bias AUCs (G-AUC)* [11]: In order to assess whether these methods contribute to reducing text biases, we take into account the prediction of labels related to bias mitigation task and validate performance using Generalized Mean of Bias AUCs [11]. This metric is becoming a de-facto metric for evaluating textual biases in the follow-up and in Kaggle competitions [11].

### 4.5 Common Hyperparameter

The common hyperparameters that we use in Dbias are a batch size of 16, 10 epochs, sequence length 512, number of labels for news is 2 (bias, non-biased). The learning rate, warm-up setups, the drop-out rate and other parameters for each module are optimized according to their best settings. For fair comparison, we tune all the other methods (baselines) to their optimal hyperparameter settings and report the best results.

We use the DistilBERT (distilbert-base-uncased) with these details: Uncased: 6-layer, 768-hidden, 12-heads, 66M parameters, and we fine-tune it on MBIC dataset. We use the RoBERTa with these details: RoBERTa-base, 12-layer, 768-hidden, 12-heads, 125M parameters along with Spacy English Transformer NER pipeline[13].

## 5 Results and Analyses

The results and analyses are discussed here.

### 5.1 Comparison between baselines and Dbias

These experiments are conducted in two-fold manner: (1) evaluation before de-biasing, and (2) evaluation after the debiasing, following the standard practice in literature [2].

In the evaluation before de-biasing, we simply take the values of protected variables to measure the DI ratio. We also detect the existence of biases in the data before de-biasing.

In the evaluation after de-biasing, we apply the bias mitigation method provided by each method on the original data to create a transformed dataset. The transformed data is assumed to be fairer because the transformation is learned as a new representation of the data using some mapping or projection functions.

Lastly, we compute all the metrics (PREC, REC, F1, ACC, DI) on the transformed dataset. We also show the overall performance of each bias mitigation technique using G-AUC, adhering to the standard evaluation style for evaluating fairness in texts [11].

In these results, we are primarily interested in the fairness metrics (DI and G-AUC), but the accuracy scores associated with the classification task are also optionally displayed to reveal how effective the bias mitigation technique is (how much bias being removed). After bias mitigation, we anticipate a lower score from the classifiers. This is due to the fact that after more biases are mitigated, it is reasonable for the classifier to miss a large number of biased terms that were previously identified.

[13] en_core_web_trf

**Table 4:** Comparison of our framework with the baseline methods.

| | | *Before de-biasing* | | | | | *After debiasing* | | | | | |
|---|---|---|---|---|---|---|---|---|---|---|---|---|
| | **Model** | **PREC** | **REC** | **F1** | **ACC** | **DI** | **PREC** | **REC** | **F1** | **ACC** | **DI** | **G-AUC** |
| *Pre* | Disparate impact remover | 0.593 | 0.549 | 0.570 | 0.587 | 0.702 | 0.532 | 0.414 | 0.466 | 0.541 | 0.804 | 0.634 |
| | Reweighing | 0.613 | 0.535 | 0.572 | 0.619 | 0.702 | 0.591 | 0.524 | 0.555 | 0.604 | 0.832 | 0.653 |
| *In* | Adversarial Debiasing | 0.624 | 0.600 | 0.612 | 0.641 | 0.702 | 0.592 | 0.587 | 0.590 | 0.610 | 0.923 | 0.679 |
| | Exponentiated Gradient Reduc. | 0.612 | 0.587 | 0.599 | 0.626 | 0.702 | 0.589 | 0.557 | 0.573 | 0.606 | 0.896 | 0.645 |
| *Post* | Calibrated Equalized Odds | 0.568 | 0.479 | 0.520 | 0.560 | 0.702 | 0.563 | 0.479 | 0.518 | 0.523 | 0.829 | 0.610 |
| | Equalized Odds | 0.498 | 0.487 | 0.492 | 0.577 | 0.702 | 0.487 | 0.488 | 0.487 | 0.505 | 0.818 | 0.598 |
| | Dbias | **0.735** | **0.784** | **0.759** | **0.776** | 0.702 | **0.690** | **0.704** | **0.697** | **0.743** | **1.012** | **0.780** |

Next, we present the results of all methods (baseline methods and Dbias) in Table 4.

***Overall results:*** The results in Table 4 show that the DI ratio of all models in the 'before de-biasing' evaluation phase is constant. This is because the DI is calculated based on the original dataset before we apply any technique to our dataset.

The DI score in the 'before de-biasing' evaluation is 0.7, which shows that the unprivileged groups receive a positive outcome less than 80% of the time than the privileged groups, which is a disparate impact violation.

Table 4 also shows the performance of bias detection using the PREC, REC, F1, and ACC scores when comparing all the methods, including ours in the 'before de-biasing' testing. We observe that the performance of our classification method is the highest. Since the classification module may not be part of the baseline method, we use the AutoML to find the best classifier to be used with each method.

In the 'after de-biasing' evaluation, we observe that the DI ratio by our method has improved a lot. A good DI value is one that is between 0.8 and 1.25, ensuring that different groups (gender, race, education, and such) are balanced [46]. Our model has a DI ratio of 1.012, which indicates we are able to mitigate biases among various groups in an appropriate and balanced manner. The G-AUC score of our approach is 78%, which is quite high in comparison with the baseline methods.

***Baseline comparisons:*** Among the baselines, the general performance of the fairness in-processing methods is better than the pre-processing methods, which is better than the post-processing methods. This is most likely due to the fact that the fairness in-processing models have explicit control over the optimization function of a model. As a result, these models can better optimize the measure of fairness during the model training.

The pre- and post-processing approaches do not change a model explicitly. This means that, in their current state, any ML library can be used for model training in pre- and post-processing approaches. However, this comes at the cost of having no direct control over the optimization function of the ML model. Thus, each technique has a tradeoff, depending on whether we need more explicit control over the method (as in in-processing) or whether we need to incorporate fairness without affecting model training (as in pre- and post-processing).

Among the baselines, the in-processing methods provide much fairer results by mitigating the biases, as shown by improved DI. In-processing methods work by incorporating one or more fairness measures into the model optimization functions in order to converge on a model parameterization that maximizes both performance and fairness [2]. Between the two in-processing baselines, we see that the performance of Adversarial Debiasing [24] is better than the Exponentiated Gradient Reduction method [25]. Adversarial Debiasing has a DI ratio of 0.923, while Exponentiated Gradient Reduction has a ratio of 0.896. In the same way, we see better G-AUC score of Adversarial Debiasing in the results.

Based on above results, the Adversarial Debiasing provides fairer results between the two in-processing techniques. The Adversarial Debiasing methods uses the adversarial training method to enforce a fairness constraint during the model optimization. The Exponentiated Gradient Reduction yields a randomized classifier with the lowest (empirical) error subject to the desired fairness constraints. This result indicates that adversarial training can be a useful technique to ensure fairness in NLP solutions.

Next comes the performance of pre-processing methods. Pre-processing methods usually change the sample distributions of protected variables or perform transformations on the data to make it less discriminatory in the training set [2], [7]. In this case, the main idea is to train a model on a 'repaired' dataset. Between the two pre-processing methods among our baselines, we see that the general performance of Reweighing [19] is better than the Disparate Impact Remover method [12]. This is shown with the better scores of Reweighing during evaluation. The DI ratio of Reweighing after de-biasing is 0.832, which is better than that of the Disparate impact remover method and also better than its own 'before debiasing' evaluation score. However, a value of 0.832 is still close to 0.8, which means that it is favoring privileged groups.

Last comes the performance of post-processing methods. Post-processing methods mainly apply transformations to model predictions to improve fairness [2]. Between the two post-processing methods, the general performance of Calibrated Equalized Odds [28] is better than the Equalized Odds method [27]. This is shown with the better scores of Calibrated Equalized Odds during both evaluation phases. The DI ratio of Calibrated Equalized Odds is 0.829, which is a better value than the other method but still not close to 1 (a value around 1 means balancing between different groups).

Compared to the baselines, our method is able to achieve the highest classification accuracy in the after de-biasing evaluation. While classification accuracy is not the goal here, the fairness is the evaluation criteria to meet. When checking the two fairness criteria, the DI ratio of our approach is close to 1, which shows that we are able to achieve a balance between unprivileged and privileged groups.

Our model's performance in terms of G-AUC is about 78%, which may not be too idealistic, however it is quite significant when considering its impact on the frequency of false positives. This is demonstrated by the model's effectiveness in lowering the bias and false-positive rate (evidenced by an increase in precision in the 'after de-biasing' testing) compared to other methods.

***Tradeoff between accuracy and fairness***: Overall, these results indicate that there is a tradeoff between accuracy and the fairness (DI ratio) measures. In the 'after de-biasing' testing, we observe an impact on classification accuracy, which is justifiable. Since the 'after debiasing' stage involves detecting bias from debiased sentences in which biased words have been replaced, it is expected that the accuracy of the bias detection will be reduced. This is obvious because after debiasing, the sentence cannot be successfully detected as being originally biased, which aligns with the previous research [2], [8].

### 5.2 Effectiveness of the Bias Detection Module

In this experiment, we evaluate the performance of our framework for the bias detection module (the first module of the Dbias). We fine-tune the bias detection module using different models and embeddings. The goal is to see which model gives us the best results for the classification task. We also compare these methods with our fine-tuned DistilBERT model.

We use the following models in this experiment. Some of these methods are traditional ML methods and some are deep neural network methods, we also use the Transformer-based methods in this experiment, which are advanced deep neural methods with self-attention.

*Logistic Regression-TFIDF Vectorization (LG - TFIDF):* We use the Logistic Regression (LG) with TfidfVectorizer[14] word embedding method. Logistic regression + TFIDF Vectorization has shown to be a good baseline method for many classifications tasks, such as hate speech detection, text classification.

*Random Forest + TFIDF Vectorization (RF-TFIDF):* We use the Random Forest (RF) classifier with TfidfVectorizer word embedding. RF + TF-IDF Vectorization are also used for text classification, sentiment analysis and related tasks.

*Gradient Boosting Machine + TFIDF Vectorization (GBM-TFIDF):* We use the Gradient Boosting Machine (GBM) with TfidfVectorizer word embedding method.

*Logistic Regression + ELMO (LG-ELMO):* We use LG with ELMO embeddings. ELMo is a contextual word embedding technique based on bi-directional LSTM.

*MultiLayer Perceptron + ELMO (MLP- ELMO):* We also use the MLP[15], a feedforward artificial neural network with ELMO embeddings, which has shown good performance for the classification tasks.

*Bert-base*: Bidirectional Encoder Representations from Transformers (BERT) is Transformer model pretrained on a large corpus of English data in a self-supervised fashion [32]. We use the bert-based uncased[16] in this work.

[14] TfidfVectorizer
[15] MLP
[16] bert-base-uncased

*RoBERTa-base:* Robustly Optimized BERT Pre-training Approach (RoBERTa) [43] optimizes the training of BERT with improved training methodology.

*DistillBERT*: We also use the DistilBERT [42], which is a small, fast, cheap and light Transformer model trained by distilling BERT base.

BERT, RoBERTa and DistilBERT have shown good performance in many classification tasks, including text classification. These are Transformer-based approaches.

The results for the evaluation of bias detection module are shown in Table 5.

**Table 5**: Effectiveness of different classification models

| Model | PREC | REC | F1 |
|---|---|---|---|
| LG-TFIDF | 0.62 | 0.61 | 0.61 |
| RF-TFIDF | 0.65 | 0.64 | 0.64 |
| GBM - TFIDF | 0.65 | 0.66 | 0.65 |
| LG- ELMO | 0.66 | 0.68 | 0.67 |
| MLP- ELMO | 0.69 | 0.67 | 0.68 |
| Bert-base | 0.72 | 0.69 | 0.70 |
| RoBERTa-base | 0.75 | 0.70 | 0.72 |
| DistilBert | **0.76** | **0.74** | **0.75** |

Overall, the results in Table 5 show the better performance of deep neural network embeddings (i.e., ELMo) compared to TFIDF vectorization when used with LG, RF and GBM. The deep neural embeddings and the deep neural methods (MLP, BERT, RoBERTa and DistilBERT) also perform better than the traditional ML methods.

There is a slight performance difference among the three classical ML approaches (LG-TFIDF, RF-TFIDF, and GBM-TFIDF), with GBM-TFIDF marginally performs better than the other two models. This is demonstrated by GBM- TFIDF's performance, which results in a 1% higher F1-score when compared to RF-TFIDF and a 4% increase when compared to LG-TFIDF.

Using ELMO embeddings, we can see that the LG model's performance has improved by about 5% compared to LG-TFIDF. This is demonstrated by LG-ELMO having a higher F1-score (0.70) than LG-TFIDF (0.65). Additionally, the LG-ELMO outperforms other traditional machine learning approaches by approximately 1-6 % better F1-scores.

The results in Table 5 also show that the deep neural baseline MLP-ELMO performs better than traditional ML baselines as well as the LG-ELMO (a mix of classical ML and deep neural network embedding model). The result shows that deep neural embeddings, in general, can outperform traditional embedding method (e.g., TFIDF) in text classification tasks (e.g., the bias detection in this work). This is probably because, deep neural embeddings can better capture the context of the words in the text in different contexts. The use of deep neural embeddings in conjunction with a model based on deep learning, such as MLP, further improves the results, as demonstrated by the better performance of MLP-ELMO model compared to ML baselines and LG-ELMO. Though the deep neural network methods perform better in many NLP tasks, the traditional ML methods are usually faster and computationally less expensive.

We also use the Transformer-based embeddings, such as from BERT, which are on dynamic word embeddings. Transformers are large encoder-decoder models that employ a sophisticated attention mechanism to process an entire sequence. The results show that Transformer-based methods outperform the other methods (ML and simple deep learning-based methods) in the bias detection task. Among the Transformer-based approaches, RoBERTa outperforms the BERT model by approximately 2% in terms of F1-score, while DistilBERT outperforms the RobBERTa model by approximately 3%.

DistilBERT is smaller, faster, and lighter than BERT and RoBERTa. When we apply DistilBERT to our dataset, it also performs significantly better than all the other models, as shown in Table 5. As a result, we choose to work with the DistilBERT in our bias detection module,

### 5.3 Effectiveness of bias recognition module

We also test the effectiveness of our bias recognition module. We compared several configurations of NER models, such as ML-based NER pipelines, Transformer-based approach, until we obtained a configuration that we consider to be the best for our goals. Next, we see the performance of different NER pipelines for the bias recognition task:

**Spacy core web small (sm) pipeline**[17] **(core-sm):** This pipeline consists of following components: token-to-vector, tagger part-of-speech tagging, parser, Sentence Recognizer, Named Entity Recognition, attribute ruler and a lemmatizer. It is an English pipeline trained on written web text (blogs, news, comments), which includes vocabulary, syntax

[17] en_core_web_sm

and entities. It is called as 'core' as it is based on traditional ML and packaged in the Spacy Core library, which use a CPU-optimized pipeline. The 'small' in this pipeline refers to the size of mode, which is 13 MB in this case.

**Spacy core web medium (md) pipeline[18] (core-md):** This is the same pipeline as core-sm but with medium size, which is 43 MB. The model size refers to different configurations, e.g., trained on more data with different parameters, numbers of iterations, vector size and such.

**Spacy core web large (lg) pipeline[19] (core-lg):** This is the Spacy core NER pipeline just like the core-sm and as core-md but with large size, which is 741 MB.

**Spacy core web transformer (trf) pipeline[20] (core-trf):** This is spacy core NER pipeline, but it has Transformer as the vectorizer. The difference between core-trf and other core pipelines is in the embedding model. The core-trf uses the RoBERTa [43], Transformer as embedding model that helps to automatically identify and extract entities from the text. The results of different NER pipelines to recognize the biases are shown in Table 6.

**Table 6:** Evaluation of different NER pipelines

| Model | PREC | REC | F1 | ACC |
|---|---|---|---|---|
| **Core-sm** | 0.59 | 0.27 | 0.37 | 0.37 |
| **Core-md** | 0.61 | 0.45 | 0.52 | 0.53 |
| **Core-lg** | 0.60 | 0.62 | 0.60 | 0.67 |
| **Core-trf** | **0.66** | **0.65** | **0.63** | **0.72** |

The results in Table 6 show that core-trf outperforms all the other NER methods in terms of precision, recall, F1-score and accuracy. This is because a Transformer-based model can identify entities and relations within the text and can generate a text representation that takes the context of each term into account. Additionally, these findings indicate that model performance in terms of accuracy metrics improves as model size increases. This is likely due to the fact that the larger model contains a greater number of parameter settings and data points, all of which affect the model's predictive performance. However, these benefits come at the expense of resource utilization, memory, CPU cycles, and latency delay. Based on these results, we choose to work with core-trf to recognize the bias-bearing words from the news articles.

[18] en_core_web_md
[19] en_core_web_lg
[20] en_core_web_trf

## 5.4 Effectiveness of masking technique

In this section, we explore the influence of masking in the MLM technique. We use different settings for the mask: we replace 5% of the input sequence with [MASK] with a probability *p* of 0.1, 0.3, 0.5, 0.8 and 1.0, as in related works [32], [47]. There is no standard masking percentage, but we use 5% based on the sentence length (around 10-40 words) of news text in our dataset. We then compare these settings with our masking technique where we only replace the bias-bearing words and we don't make use of any random probability as in typical MLM tasks [32], [47]. The goal of this experiment is to see if using different masking techniques (with different probabilities) affects the final output. The results of are shown in Figure 8.

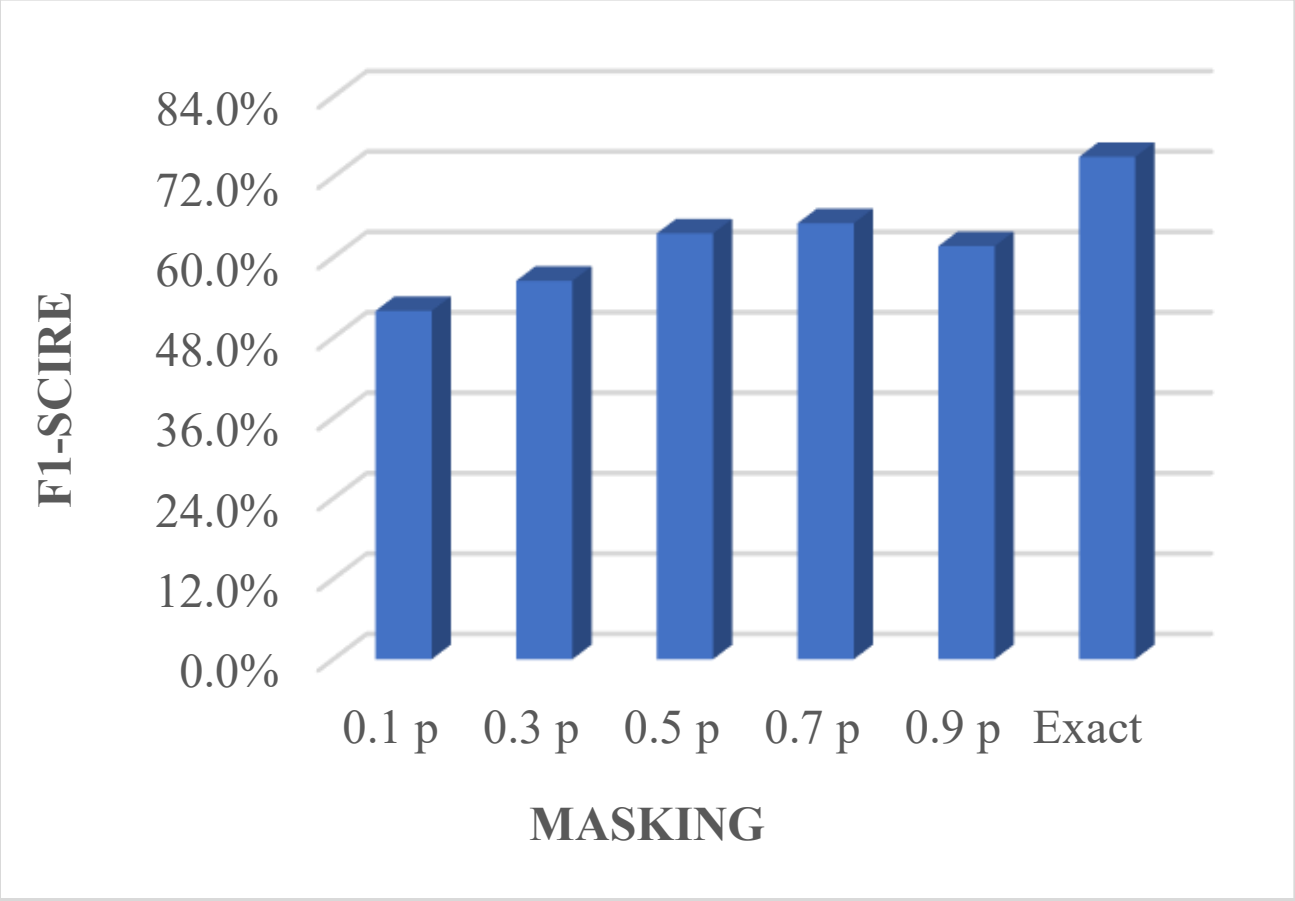

**Figure 8**: probability *p* vs exact bias mask

The results in Figure 8 show that the performance of Dbias improves when probability *p* increases from 0.1 till 0.7. When *p* is too small, the model perhaps tends to overfit, since our model has many parameters. Thus, the performance is not optimal. However, when *p* is too large i.e., 0.9, the performance of model starts to decline. In contrast, our exact masking of biased words shows the best performance.

The general conclusion that we can draw from these results is that if we replace the input sequence with [MASK] using varying probabilities, the model performance may be affected, as the model learns to detect only the masked word, rather than actual word representations. As our use case in this work is masking the biased words only from the text, an exact match can give us better results.

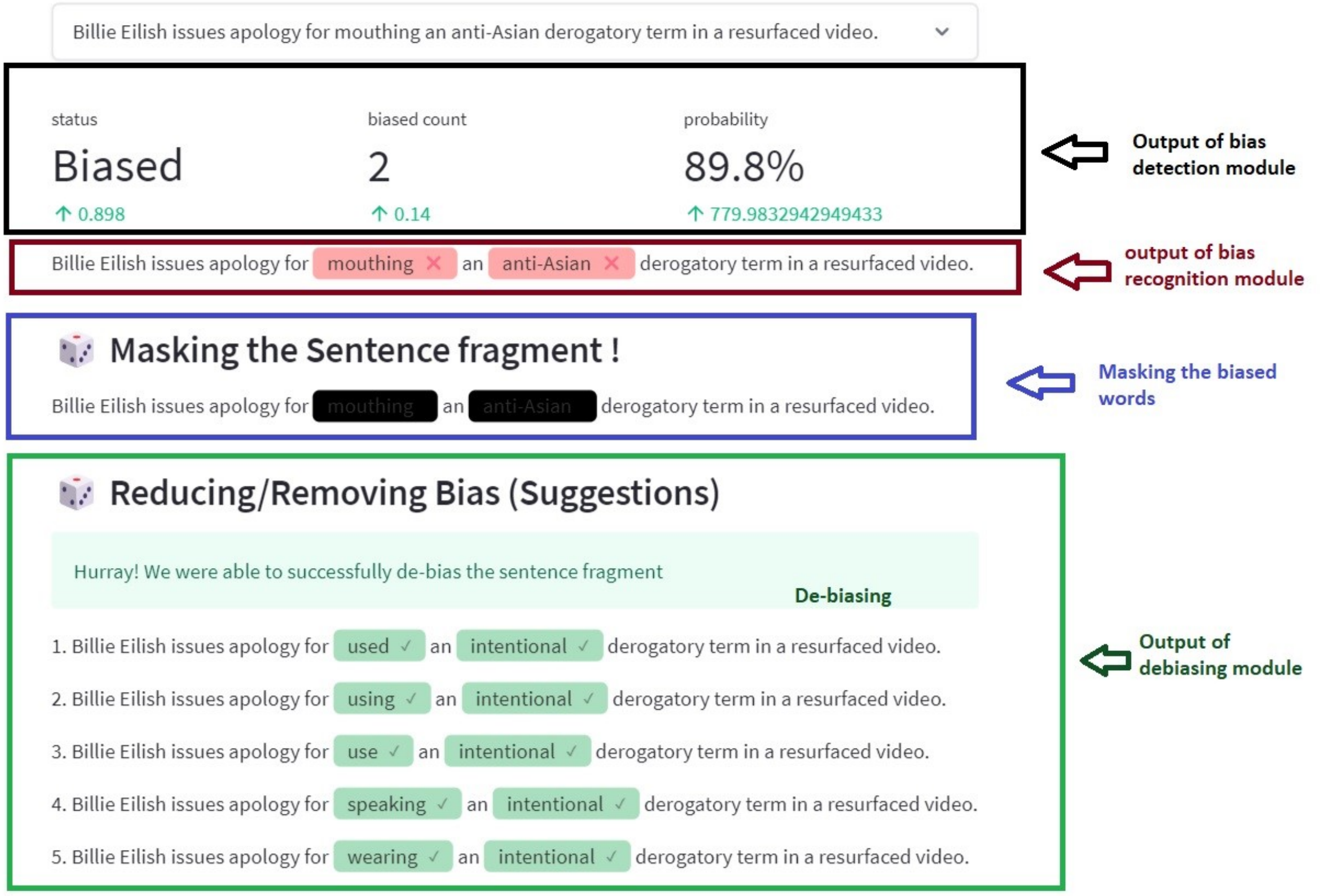

**Figure 9: Working of Dbias**

## 5.5 Working of Dbias model

We release our Dbias package [21] that is used to detect and mitigate biases in NLP tasks. The model tasks are summarized as shown in Table 7:

**Table 7:** Dbias tasks

| Feature | Output |
|---|---|
| **Text Debiasing** | Returns debiased news recommendations with bias probability |
| **Bias Classification** | Classifies whether a news article is biased or not with probability |
| **Bias Words Recognition** | Extract Biased words or phrases from the news fragment |
| **Bias masking** | Returns the news fragment with biased words masked out |

The model can be installed using the commands:

- `pip install Dbias`
- `pip install` [https://huggingface.co/d4data/en_pipeline/resolve/main/en_pipeline-any-py3-none-any.whl](https://huggingface.co/d4data/en_pipeline/resolve/main/en_pipeline-any-py3-none-any.whl)

The input to the model can be any sentences that may contain biased words and we get the de-biased output. We show the working of Dbias in Figure 9.

[21] https://pypi.org/project/Dbias/

As illustrated in Figure 9, given a news article or any text that may contain biased words, our model can determine whether or not the text is biased. This is made possible by the first module: bias detection module. The output is then forwarded to the next module, namely the bias recognition module, which identifies bias-bearing words. The text with identified biased words is then sent to the de-biasing module, which masks the biased words and makes suggestions for new words to replace them. The final output is a set of non-biased or at least minimally biased sentences for each input sentence.

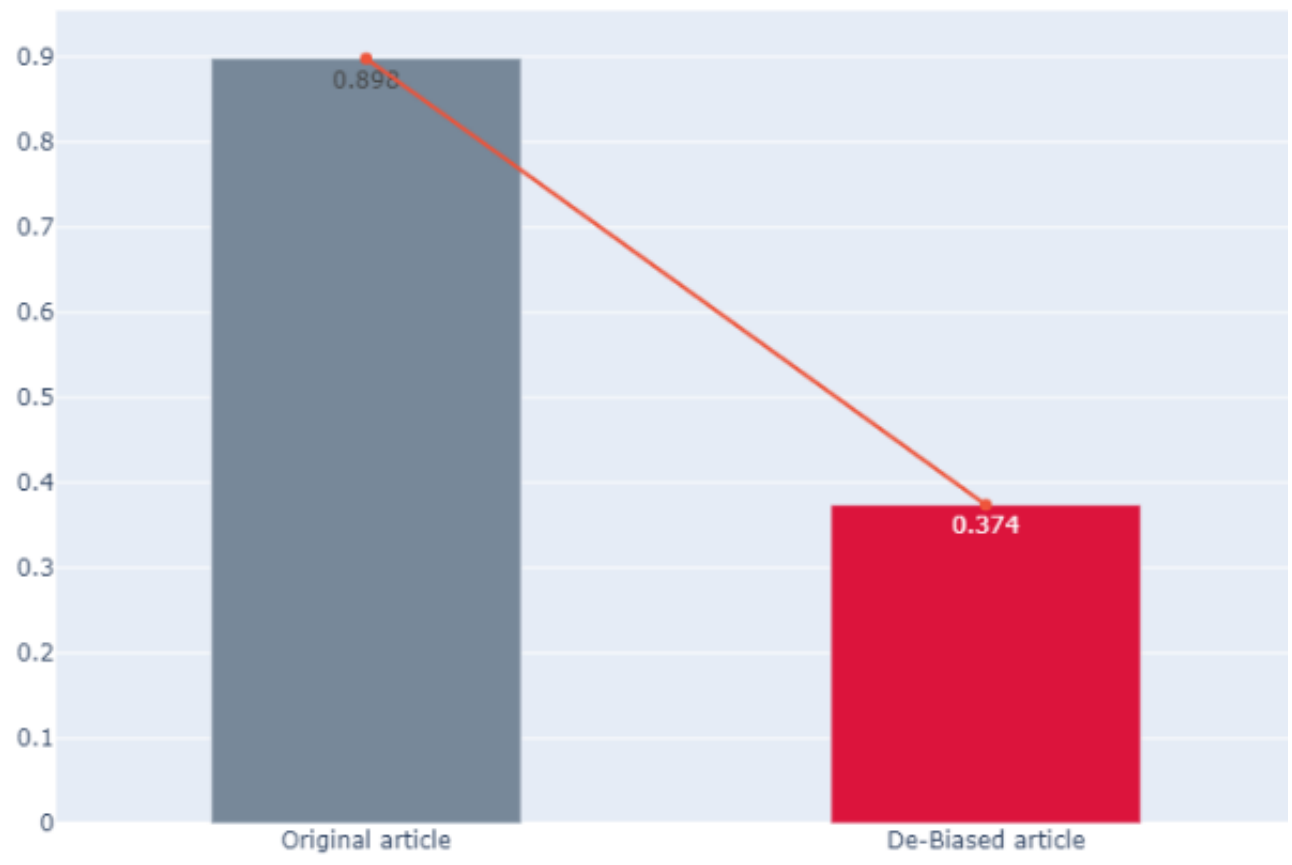

**Figure 10:** Bias probability in a news article

We also show the bias probability that is achieved from an original news article to a de-biased article in Figure 10. We perform a forward pass to compute logits and apply softmax to calculate probabilities. A lower score here means that less bias exists and therefore detected with smaller probabilities.

As shown in Figure 10, our Dbias framework can reduce the biasness of a news article by about 50%.

## 6 Discussion

Fairness in ML and AI is a relatively new but rapidly growing field of study in areas such as information retrieval, recommender systems, and so on. The broader literature on fairness in ML is a crucial starting point, therefore, we undertook this research with great care to avoid pitfalls that might result in overbroad or ungeneralizable claims.

The research [4] shows that recommending unbiased news to users broadens their perspective on the news and on society as a whole. By repeatedly exposing users to news containing biased language, their perceptions of certain demographic groups or the news story itself may be altered. Through this package, we make an effort to provide news that is free of bias or with less bias. While this may not be the ultimate solution, we provide a means to contribute to the dissemination of news that is more genuine and freer of societal and other biases.

Comparison with the state-of-the-art methods: The state-of-the-art fairness algorithms like FIXOUT [48] and its advanced version [49] are also used to reduce biases in the tabular and textual data. These models define an ensemble classifier that can use any explanation method (based on feature importance), such as Shapley Additive Explanations (SHAP) and/or Local interpretable model-agnostic explanations (LIME), to explain the predictions of the underlying models (classifier). This helps to assess model's reliance on sensitive features. Our method can also detect the sensitive attributes (selection of protected groups) based on the biased words in text. However, our work differs from [48], [49] that we detect the biases and then de-bias those words through MLM task. We think by including the explanation component, such as SHAP OR LIME in the proposed method, we would be able to explain the predictions of the classification process.

Our focus in this work is on mitigating biases in the textual data, which is different from the detecting and correcting biases in the numeric data [50], [51]. While, we keep our focus on the NLP fairness, we take the direction from this work [50] to improve the detection tasks by converting a binary classification problem to multiple classification problem. we believe this would also help to alleviate the negative impact of the large span of the training labels.

Some other research directions are to extend the current study to mitigate sentimental biases in the texts [52]. This study [52] proposed a new attention mechanism, called polar attention, to mitigate sentimental biases. We think, including the polarity of sentiments, a model can reduce the extent of sentimental bias for neutral words while truly attending the polar words and to reduce their impact in the texts.

As with any research study, there are limitations. In order to help understand and generalize the results, we have taken extra care to disclose the limitations of the data, metrics, and methods used. There are some limitations that we want to discuss here, and we also discuss some future recommendations.

***Different definitions of biases and fairness****:* There is currently no universally accepted definition of what constitutes bias and fairness. The findings drawn about one bias cannot be generalized to other biases. Additionally, biases manifest in a variety of ways (e.g., the bias definition of gender cannot be applied to ethnicity or social status). While more standard definitions of biases and fairness may be discovered in the future, we must first investigate a wide variety of biases in different applications to determine the fairness of data and algorithms.

***Biases evolve over time*****:** While much of the research in this field considers a small number of biases (social, demographic, and so on), we emphasize that there are many other types of biases that are overlooked in the research. For instance, numerous biases have arisen in recent years as a result of COVID-19 or U.S ELECTIONS 2016 and 2020 [53]. Taking this limitation into account, we develop a system capable of dynamically removing biases from data. This means that an initially fair system may become unfair over time if users respond in a biased manner, or that it may evolve toward a fairer system if users respond positively to recommendations that improve overall fairness. We may need to investigate additional biases that evolve over time and to consider them in the Dbias pipeline.

***Fairness metrics****:* There are many fairness metrics, for example, statistical parity, predictive parity, calibration, pairwise fairness and others, as discussed in the literature [2], [6], [10]. There is still much work left to do to understand how best to apply and interpret these fairness metrics in our study. We only apply a few fairness metrics, like DI and G-AUC. We may need to explore other metrics and check how they affect the performance, or maybe we need to do some modifications/justifications of the pipeline to optimize for the other metrics.

***Use of appropriate data:*** Data collection is a significant challenge for fairness research because it frequently requires sensitive data that cannot be collected via standard information retrieval or recommender system data sets. We use a manually annotated news dataset in this study to identify bias-bearing words. We encourage researchers to use free texts to identify and mitigate more biases in the text.

***Transfer learning***: In our Dbias modules, we used the transfer learning technique. We recognize that transfer learning may introduce additional biases. However, we find that carefully fine-tuning the models on the appropriate data can be useful. For example, we fine-tune our modules on news data and then these models help us to detect and mitigate various biases from the news data.

***Scarcity of labelled biased data***: One of the research limitations is the scarcity of labelled biased data. So far, we have relied on the MBIC dataset to assist us in detecting and mitigating bias, which has been validated through extensive experiments. However, we would like to acquire more labelled data in the future to train our package. We also plan to consider annotating the data using unsupervised text classification, such as with zero-shot approach [54].

***Biases in different contexts***: In this work, we replace biased words with non-biased words, thereby modifying the words. Our analysis of the results indicates that changing the words does not affect the news content or the overall opinion. Multiple examples are provided in our package release (https://pypi.org/project/Dbias/). While this is an attempt to preserve the meanings of the actual news while removing only biased words, we believe that additional testing may be necessary in the future.

While contextual patterns were considered during the identification and mitigation of biases, additional linguistic and stylistic features are required to capture the overall aspect of biases in the entire news article. This can be a potential future direction. We also intend to include an explanation for articles that our system deems biased. In addition, it would be beneficial in the future to flag an article as biased if it is labeled as biased and contains biased language.

***Crowdsourcing and biases:*** We recognize that crowdsourced datasets frequently contain substantial social biases, such as gender or racial preferences and prejudices. The algorithms trained on these datasets may produce similarly biased decisions. One future direction in this regard is to measure crowd workers' biases based on counterfactual fairness. Counterfactual fairness [55] refers to the intuition that a decision is fair to an individual if it is the same in (a) the real world and (b) a counterfactual world in which the individual belongs to a different demographic group. We also suggest that the providers of datasets in this area of study make the annotation process more transparent. To provide greater openness in the debiasing process, we plan to provide explanations for each outcome on our end.

***Future perspectives***: There is still much work to be done toward achieving fairness in ML, and we hope that others in the research community will continue to contribute their own approaches to fairness and bias checking, mitigation, and explanation to the toolkit. This package is developed to de-bias news articles, anyone can use it to train on other types of data, such as journalism, hiring applications, prison sentencing or health science, and then use it to de-bias that data. As a result, one future direction is to extend the toolkit's application to additional datasets, such as toxic comments datasets released by the Conversation AI[22] team, a research initiative founded by Jigsaw and Google.

## 7 Conclusion

We build Dbias, a pipeline for fair ML, which is composed of three main modules: bias detection, bias recognition, and de-biasing. We develop a Transformer-based model to detect biased news using labelled news data; we develop an NER recognition model based on the Transformer architecture to identify biased words in biased news articles; and we use the MLM technique to replace the biased words in the text with neutral words. We compare Dbias performance to state-of-the-art fairness methods. Additionally, we evaluate the efficacy of individual components of Dbias. We release Dbias as a freely downloadable package for the users and practitioners. This research serves as a forum for researchers interested in de-biasing the text. The package can be used by developers to detect and mitigate bias.

[22] https://conversationai.github.io/

## Compliance with Ethical Standards

**Conflict of Interest**: On behalf of all authors, the corresponding author states that there is no conflict of interest.